\documentclass[floatfix,aps,showpacs,twocolumn,nofootinbib,preprintnumbers]{revtex4}
\usepackage{graphicx}
\usepackage{epsfig}
\usepackage{bm}
\usepackage{longtable}

\def\lsim{~\rlap{$<$}{\lower 1.0ex\hbox{$\sim$}}\;}
\def\gsim{~\rlap{$>$}{\lower 1.0ex\hbox{$\sim$}}\;}
\newcommand{\met}{{\rm\not\!\!E}_{T}}

\begin{document}

\preprint{IFT-P.0xx/2007}

\title{Technicolor Contribution to 
Lepton + Photon + $\bf\met$ Events at the Tevatron}

\author{Alfonso R.~Zerwekh}
\email{alfonsozerwekh@uach.cl}
\affiliation{Instituto de F{\'{\i}}sica, Facultad de Ciencias,
Universidad Austral de Chile, Casilla 567, Valdivia, Chile}

\author{Claudio O.~Dib} 
\email{claudio.dib@usm.cl}
\affiliation{Department of Physics, Universidad T\'ecnica Federico Santa Mar{\'{\i}}a, Valpara{\'{\i}}so, Chile}

\author{Rogerio Rosenfeld} 
\email{rosenfel@ift.unesp.br}
\affiliation{Instituto de F\'{\i}sica Te\'orica - S\~{a}o Paulo State University, Rua
Pamplona, 145, 01405-900, S\~{a}o Paulo, SP, Brazil}

\date{\today}

\begin{abstract}
Events with one lepton, one photon and missing energy are the subject of
recent searches at the Fermilab Tevatron. We compute possible contributions to
these types of events from the process  $p \bar{p} \rightarrow  
\gamma l \nu_{l} \nu_\tau  \bar{\nu}_\tau$, where $l=e,\mu$ in the
context of a Low Scale Technicolor Model.   
We find that with somewhat tighter cuts than the ones used in the CDF
search, it could be possible 
 to either confirm or exclude this model in a small region of its parameter
space. 
\end{abstract}


\maketitle


The standard model of the electroweak interactions has been extremely successful
in describing all the high energy accelerator data collected so far \cite{EWWG}.
However, there are several reasons to believe that the standard model is incomplete, 
such as the existence of non-baryonic dark matter and non-zero neutrino masses. 
Furthermore, the infamous problems of triviality and naturalness related to the 
scalar Higgs sector of the theory point to the possibility
that the standard model is an effective theory valid up to an as yet unknown 
high energy scale $\Lambda$.
Extensions of the standard model such as supersymmetric models, models with extra 
dimensions (universal or otherwise) and models with dynamical electroweak symmetry  
breaking (DEWSB) are the main contenders for describing Nature at
energies above $\Lambda$.   
With the starting of the Large Hadron Collider at CERN one will be hopefully able 
to figure out in a few years what the completion of the standard model really is at 
around the TeV scale, if there is one. 

Meanwhile, the Tevatron is accumulating data at $\sqrt{s} = 1.96$ TeV
and it is of foremost importance to place constraints on these contenders using
what is available now.
In particular, motivated by signatures of new physics beyond the standard model, 
the CDF collaboration has recently performed a search for inclusive events with 
one lepton and one photon \cite{CDFprl06}.

In this Brief Report, we will focus on the interesting channel 
$p \bar{p} \rightarrow l + \gamma +  \met$, with $l=e,\mu$ where a $2.7 \sigma$ excess from
the standard model prediction was reported earlier using Run I data \cite{CDFprl02}.
We will compute contributions to this process from models with DEWSB 
and find bounds on this model from recent Run II data.

Models with DEWSB involve new interactions that become strong near the TeV scale
\cite{HillSimmons}. The first models were inspired by a scaled-up version of QCD, with a
new interaction called technicolor (TC) that causes new fermions in the 
fundamental representation of an $SU(N_{TC})$, called technifermions (T), to condense and 
break both a global chiral and the electroweak symmetries \cite{first}. 
Of the resulting Nambu-Goldstone bosons, called 
technipions ($\Pi_T$), three are ``eaten" by the electroweak gauge bosons, which gain
a longitudinal component and hence a mass term. No fundamental scalar fields are present.
The correct gauge boson masses are obtained if one requires that the technipion decay 
constant $F_T$ is fixed at  $F_T = v = 246$ GeV. 
As it happens in QCD, the strong TC interaction is also responsible for the existence
of resonances in the scattering of technipions. In consonance with the QCD analogy, 
the lightest vector resonances are called techni-rho ($\rho_T$) and 
techni-omega ($\omega_T$). Naively, they would be expected to have masses around $4 \pi F_T$.

These simple models become more baroque when one considers mechanisms to generate mass for 
the standard model fermions. A further interaction called extended technicolor (ETC),
usually modelled via a broken gauged flavor symmetry, is introduced \cite{ETC}.
The massive ETC gauge bosons communicate the DEWSB to the standard model fermions, 
generating masses of the order of $\langle \bar{T} T \rangle/M_{ETC}^2$.
A difficulty immediately arises in the top sector: a very low ETC scale seems to be 
required in order to generate a heavy top quark mass.
The combination of a low ETC scale with the large isospin violation necessary to 
explain the top-bottom mass difference proves fatal: precision electroweak measurements
rule out a simple QCD-like TC model with a naive ETC mechanism \cite{Isospin}.
However, further developments based on the so-called walking technicolor 
(where the TC coupling runs slowly between $M_{ETC}$ and $F_T$) \cite{walking}, which may or may not
invoke technifermions in higher representations of the TC group combined with new precision 
measurements have shown that it is possible to reconcile more sophisticated models with 
current experimental data \cite{NewModels} and even possibly with unification ideas \cite{Unification}. 
The walking property enhances both the 
standard fermions masses and, more importantly to this work, the technipion masses.


We will be interested in a variation of the basic technicolor models with far reaching 
phenomenological consequences. It concerns the possibility of lowering the TC scale 
$F_T$.
This class of models, usually called Low Scale TC (LSTC) models, 
arise in cases where sectors with different condensation 
scales are present \cite{multi}. 
The vector resonances associated with the lowest scale can be light and 
hence accessible at the Tevatron.

The phenomenology of LSTC has been extensively studied in different 
machines \cite{HillSimmons}. In particular, the resonant associated production of 
a technipion with a gauge boson via a techni-rho or techni-omega, 
$p \bar{p} \rightarrow \rho_T^{\pm} \rightarrow W_L^{\pm} \Pi_T^0$ and
$p \bar{p} \rightarrow \omega_T \rightarrow \gamma \Pi_T^0$ 
with the subsequent
decay $\Pi_T^0 \rightarrow b \bar{b}$ was analyzed in detail \cite{bb}.
The importance of the radiative decays $ \rho_T, \omega_T \rightarrow \gamma \Pi_T^0$
was emphasized in \cite{straw}.
We performed a study of the rarer but cleaner three-photon process 
$p \bar{p} \rightarrow \omega_T, \rho_T^0 \rightarrow \gamma \Pi_T^0 \rightarrow 
\gamma \gamma \gamma$ \cite{3gamma}.

In this Brief Report we will extend the analysis of \cite{3gamma}
to study the process $p \bar{p} \rightarrow  \rho_T^{\pm} 
\rightarrow \gamma \Pi_T^{\pm}$ with the subsequent decay 
$ \Pi_T^{\pm} \rightarrow \tau \nu_\tau \rightarrow l \nu_{l} \nu_\tau 
\bar{\nu}_\tau$.   

For definiteness we will at first adopt a techni-rho mass in the range 
$M_{\rho_T} = 210$ -- $300$ GeV and fix $M_{\Pi_T} = 110$ GeV.
Hence the main decay modes of the techni-rho are
$\rho_T^{\pm} \rightarrow \Pi_T^{\pm}  \Pi_T^0, W^{\pm} \Pi_T^0, Z  \Pi_T^{\pm}$ and $\gamma
\Pi_T^{\pm}$. 
In particular, the amplitude for the process that is relevant for us
can be written as \cite{straw,tomega}
\begin{eqnarray}
{\cal M}(\rho_T^{\pm}(q) \rightarrow \gamma(p_1)  \Pi_T^{\pm}(p_2))=&
  & \nonumber\\ \frac{(Q_U+Q_D )e  \cos \chi}{M_V}
 \epsilon^{\mu \nu \lambda \sigma} \varepsilon_\mu(q) 
\varepsilon_\nu^\ast(p_1) q_\lambda p_{1 \sigma}& & 
\end{eqnarray}
where $\chi$ is a mixing angle between isospin eigenstates and mass
eigenstates in the 
technipion sector, $Q_U$
($Q_D=Q_U-1$) is the charge of the techniquark up and $M_V$ is a
typical TC mass scale. We will adopt $\sin \chi = 1/3$, $Q_U=4/3$ 
and $M_V = 100$ and $200$ GeV.
With these parameters, the total techni-rho width was calculated using
Pythia \cite{Sjostrand:2006za}.
%

The charged technipion coupling to fermions is proportional to their masses.
Moreover, we assume here that the coupling is also proportional to 
Cabibbo-Kobayashi-Maskawa (CKM) mixing angles, which is reasonable if techniquarks are 
weak isodoublets. 
In our case we use $BR(\Pi_T^{\pm} \rightarrow \tau \nu_\tau) = 25 \%$,
$BR(\Pi_T^{\pm} \rightarrow c s) = 75 \%$ (the decay into $b c$ is CKM suppressed).

The production cross section can be estimated by using a generalized vector 
meson dominance argument \cite{vmd} where there is a $W-\rho_T$ mixing or equivalently 
by diagonalization of the $W-\rho_T$ mass matrix \cite{Alfonso}.
In both cases the amplitude involves a mixing
constant given by $g_{W-\rho_T} =M_{\rho_T}^2 g/(2 g_T)$, where $g$ is the $SU(2)_L$
electroweak gauge coupling and $g_T$ is the techni-rho coupling to two technipions. 
We will fix $g_T = 5.3 $, which arises from a simple QCD scaling.

As a simple figure of merit, the cross section for $p \bar{p} \rightarrow  \rho_T^{\pm} 
\rightarrow \gamma \Pi_T^{\pm}$ (for $M_{\rho_T}= 210$ GeV and $M_V = 200$ GeV)  
with a simple cut in the photon 
transverse momentum, $p_T(\gamma) > 20$ GeV, is around 1.4 pb at the Tevatron, 
whereas the background $p \bar{p} \rightarrow  \gamma W^{\pm}$ cross section with
the same cut is around $5.6$ pb. Including the appropriate branching ratios, 
it follows that  the cross section for $p \bar{p} 
\rightarrow \gamma \Pi_T^{\pm} \rightarrow \gamma l \nu_{l} \nu_\tau 
\bar{\nu}_\tau$ is around  0.12 pb compared to 1.3 pb for
$p \bar{p} \rightarrow \gamma W^{\pm} \rightarrow \gamma l + \met$.


We implemented LSTC as a CompHEP \cite{CompHEP} model using the
diagonalization of the mass-matrix procedure. We used CompHEP for
generating events for $p \bar{p} \rightarrow  \rho_T^{\pm} 
\rightarrow \gamma \Pi_T^{\pm}$. Subsequently, they were processed by
a FORTRAN code we wrote in order to generate the $\Pi_T^{\pm}$ decay
products. The dominant background from  $p \bar{p} \rightarrow  \gamma W^{\pm} $  
was generated using the same method. We checked that the background 
from $ W \rightarrow \tau \nu_\tau $ is small and hence was not included.

\begin{figure}[htb]
\vspace*{-1mm}
\centering
\includegraphics[height=6cm,width=4.2cm]{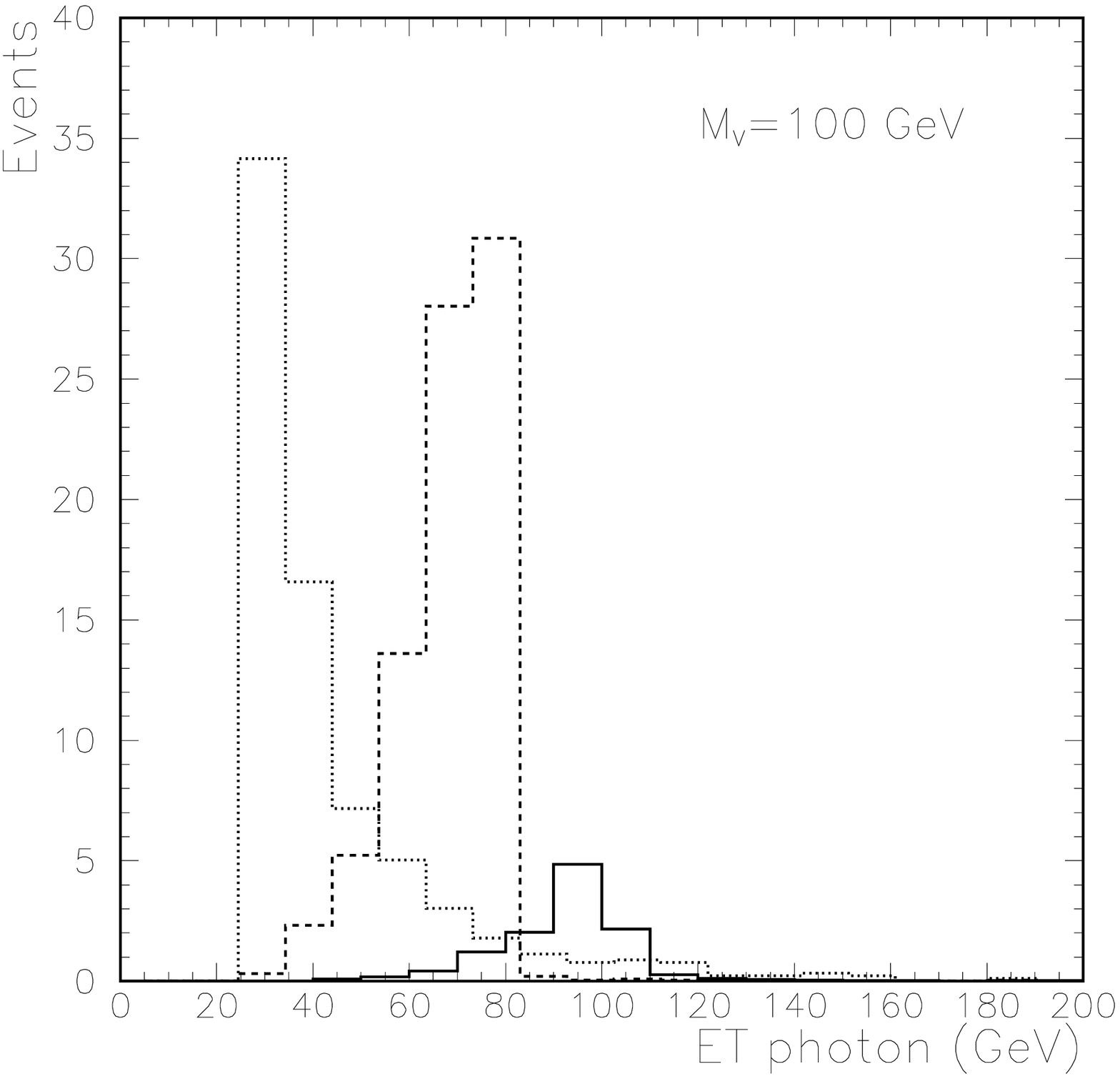}
\includegraphics[height=6cm,width=4.2cm]{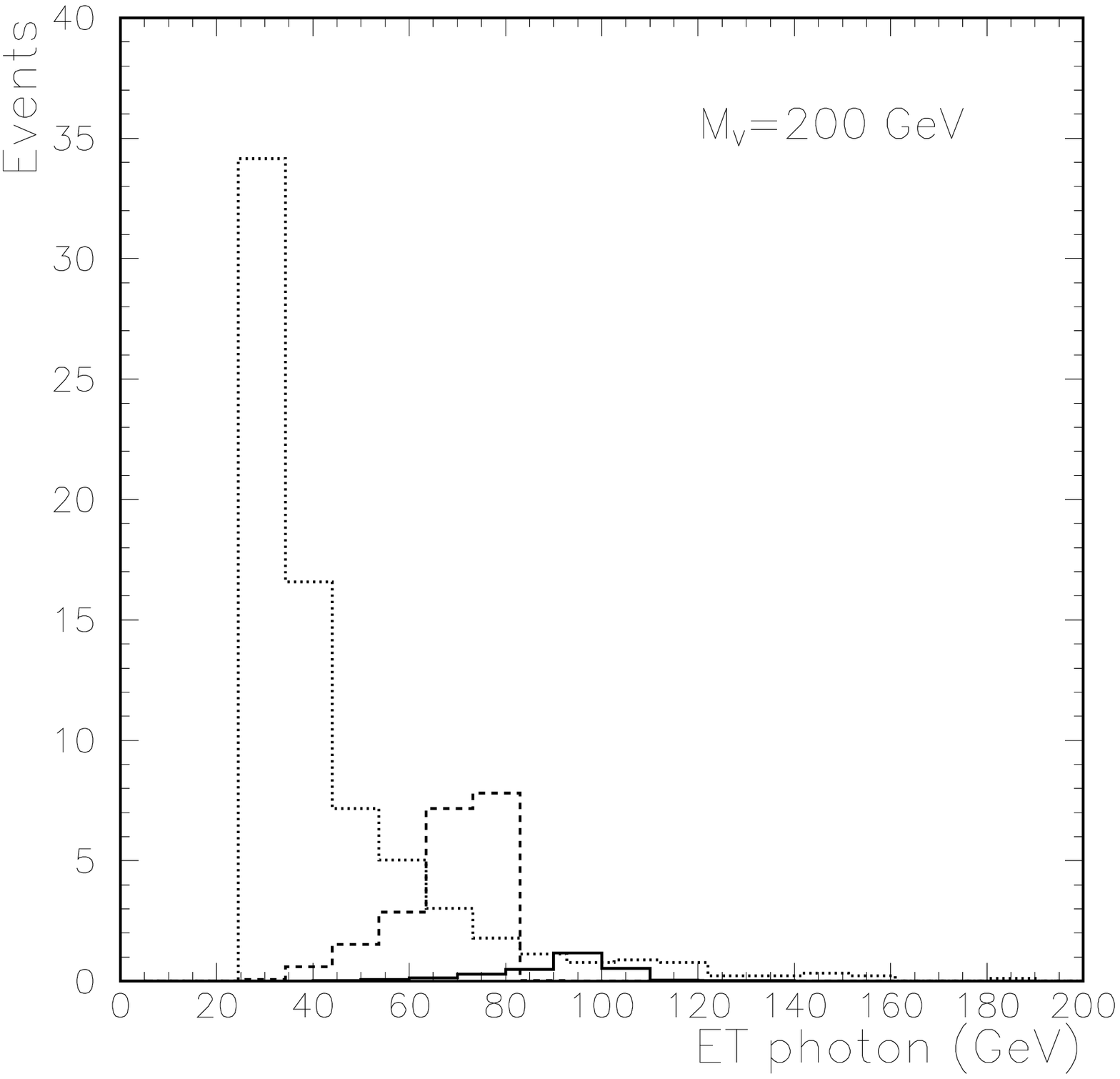}
 \caption{\label{PT}
Signal for $M_{\rho_T} = 210$ GeV (dashed line) and $M_{\rho_T} = 250$ GeV 
(solid line) for $M_V = 100$ GeV (left figure) and  $M_V = 200$ GeV (right figure) compared to the 
SM background (dotted histogram) 
for the photon $E_T$ distribution assuming a $1$ fb$^{-1}$ integrated luminosity with cuts described in the text. }
\end{figure}

In Fig. \ref{PT} we illustrate the photon transverse energy distribution for the 
signal and background for 
$p \bar{p} \rightarrow \gamma l^{\pm}  + \met$ for the Tevatron with 
an integrated luminosity of $1$ fb$^{-1}$, where $l=e,\mu$.  We choose to show only two values of the 
techni-rho mass, namely  $M_{\rho_T} = 210$ and $250$ GeV with TC scale $M_V = 100$ 
and $200$ GeV. We used the same cuts as in the CDF paper \cite{CDFprl06}, namely
$|\eta_{l,\gamma}| <1$ and $E_T^{l,\gamma}, \met > 25$ GeV.
The peak in the distribution is due to the 2-body resonant kinematics of the underlying
process.
We immediately notice that an improvement in the signal-to-background ratio
can be gained if a tighter cut $E_T^\gamma > 50$ GeV is imposed.

\begin{figure}[htb]
\vspace*{-1mm}
\includegraphics[height=6cm,width=4.2cm]{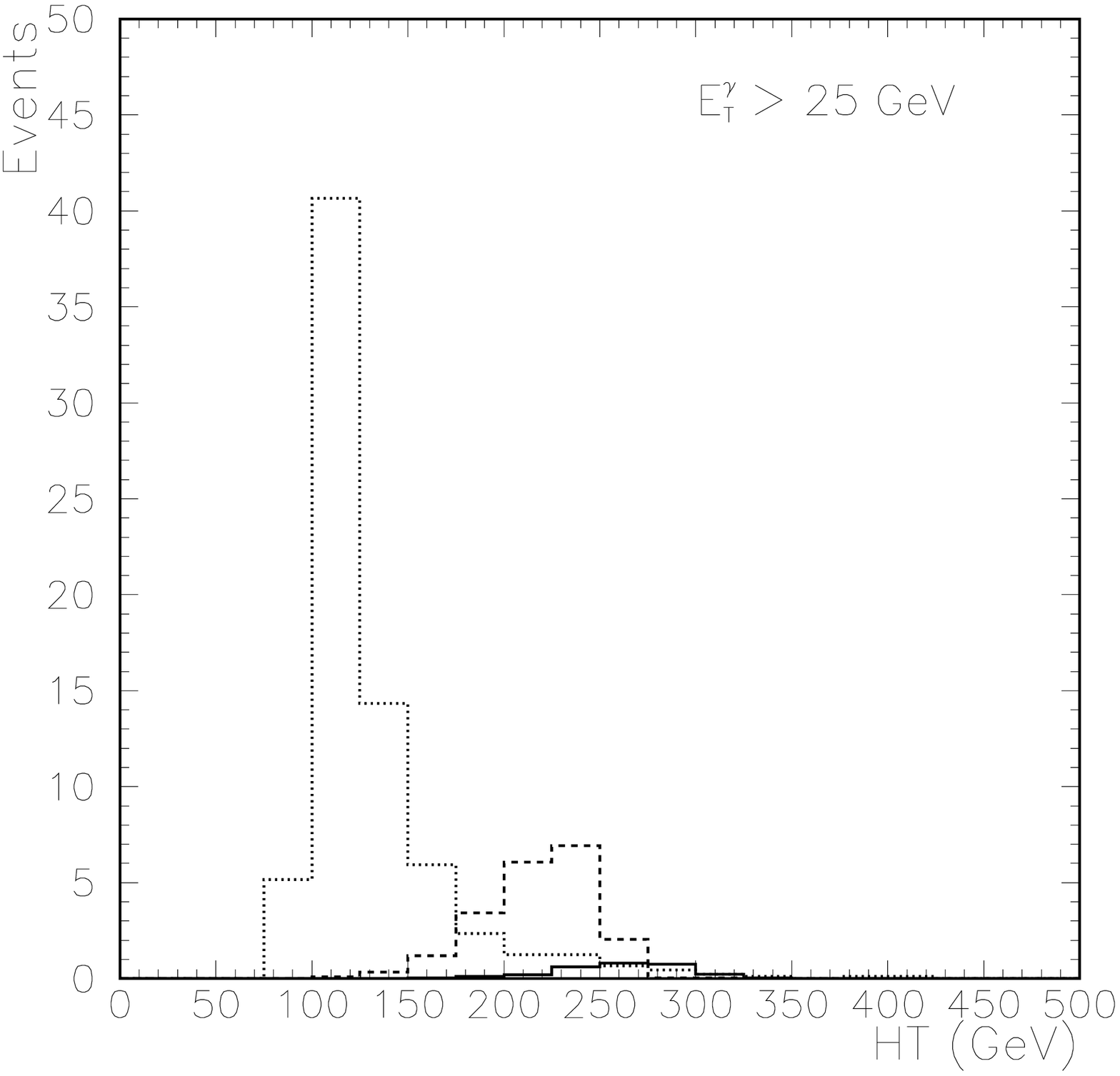}
\includegraphics[height=6cm,width=4.2cm]{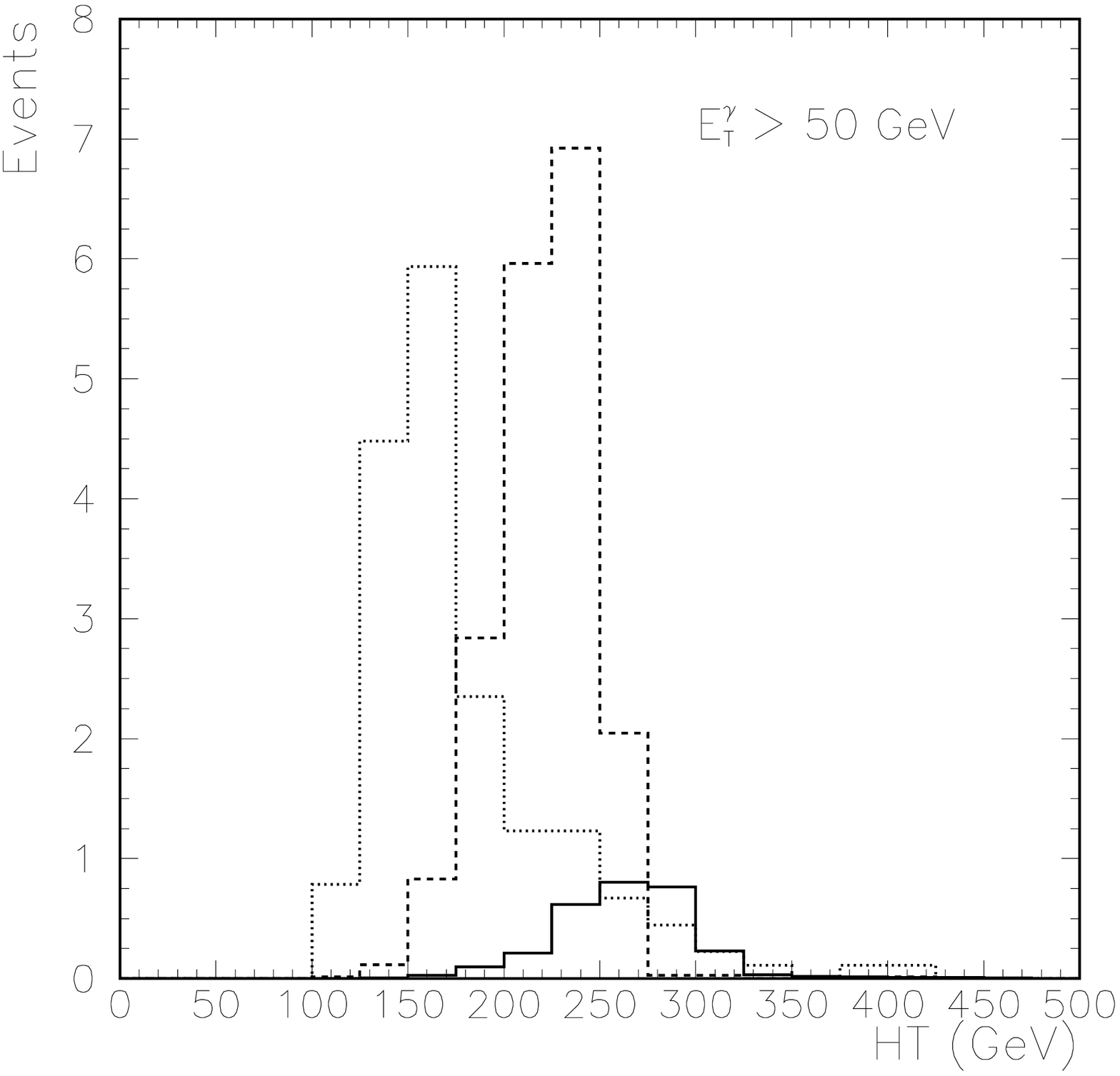}
 \caption{\label{HT}
Signal for $M_{\rho_T} = 210$ GeV (dashed line) and $M_{\rho_T} = 250$ GeV 
(solid line) for $M_V = 200$ GeV compared to the SM background (dotted line) 
for the $H_T$ distribution assuming a   $1$ fb$^{-1}$ integrated luminosity with cut
 $E_T^\gamma > 25$ GeV (left figure) and $E_T^\gamma > 50$ GeV (right figure).   }
\end{figure}

In Fig. \ref{HT} we show  the distribution of the 
variable $H_T$, defined as the total transverse energy of the event, including $\bf\met$, for
the usual CDF cut $E_T^\gamma > 25$ GeV and for a tighter $E_T^\gamma > 50$ GeV cut.
Again a tighter cut on the photon transverse energy results in a better significance of the
signal at the expense of a reduced number of events.

\begin{figure}[htb]
\vspace*{-1mm}
\includegraphics[height=6cm,width=9.0cm]{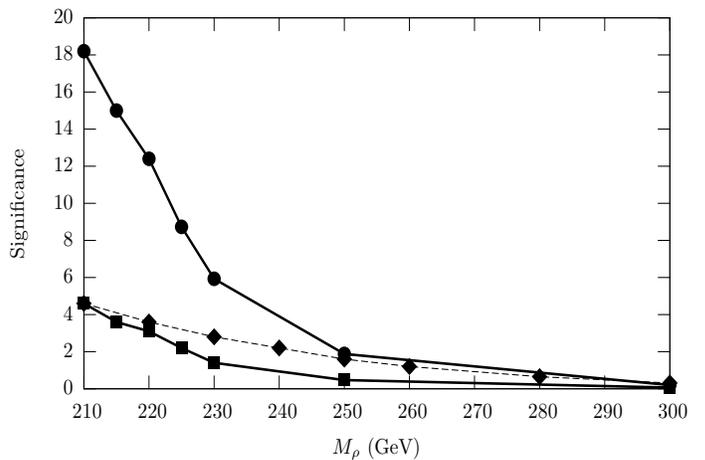}
 \caption{\label{sig}
Significance for the technicolor signal as a function of  $M_{\rho_T}$ 
for  $M_V = 100$ GeV (dots) and  $M_V = 200$ GeV (squares) for fixed $M_{\Pi_T}= 110$ GeV and 
for fixed $2 M_{\Pi_T} - M_{\rho_T}  = 10$ GeV for $M_V = 200$ GeV (diamonds).
  }
\end{figure}

The significance S can be estimated from a simple analysis involving the number of signal and background 
events in bins $i$
of the $H_T$ distribution with a tighter $E_T^\gamma > 50$ GeV, assuming Poisson statistics: 
\begin{equation}
S =  \frac{\sum_i N^{(i)}_{\mbox{signal}}}{\sqrt{\sum_i N^{(i)}_{\mbox{back}}}}
\end{equation}  
and it is shown in Fig. \ref{sig} as a function of the techni-rho mass for 
$M_V = 100$ and $200$ GeV.  
Notice that the $\rho_T \rightarrow \Pi_T \Pi_T$ channel is open in most of the technirho mass range plotted, 
as we fixed $M_{\Pi_T} = 110$ GeV (solid lines).
Since there is a rapid drop in the significance due to the opening of the two-technipion decay channel
at $M_{\rho_T} = 220$ GeV in this case, we have also studied the
case with a fixed mass difference $ 2 M_{\Pi_T} - M_{\rho_T}  = 10$ GeV, in such a way that this 
channel is always closed (dashed line).   
We find that for a TC scale as low as $M_V = 100$ GeV
a techni-rho mass below roughly  $M_{\rho_T} = 250$ GeV is excluded at the 3 $\sigma$ level.
However, for $M_V = 200$ GeV the process is suppressed but a 3 $\sigma$  could be seen for
$M_{\rho_T} < 220$ GeV. For the softer CDF cut no signal would be observed.


In sumary, we studied in this Brief Report the contribution of a Low Scale Technicolor Model
to the process 
 $p \bar{p} \rightarrow  
\gamma l \nu_l \nu_\tau  \bar{\nu}_\tau$, which falls in the class of events
recently searched for at the Tevatron. We find that with a tighter cut in the photon transverse
energy, it would be possible to either confirm or exclude this model in a small region of parameter
space. 

We would like to stress that our simulations are not fully realistic since they do not take into account
the characteristics of the detector. However, we expect that the
smearing of the final momenta be small since we have only photons and
leptons in the final states.
Our goal in this study is just to point out that the model adopted here can in fact
contribute to the production of events with one lepton, one photon and missing energy and hence should
be considered in more detailed experimental analysis.

Our analysis could be expanded in many ways if one considers hadronic final states.
In particular, the dominant $\Pi_T$ decay mode in $c \bar{s}$ may not be hopeless if a charm tagging can
be implemented in the dijet mass distribution.
Also, the case considered here of $\tau$ leptons in the final state, a signature of technicolor models,
could be better explored by using their hadronic decay modes as well.

\section*{Acknowledgments}
We would like to thank Henry Frisch for bringing to our
attention the recent searches conducted by CDF
that led to this paper and for useful comments. We also thank Ken Lane for 
a careful reading and for the questions that led to an improvement of this work. 
The work of ARZ is partially supported by grant DID-UACH S-2006-28 , COD
is partially supported by  Fondecyt, Chile, grant 1030254 and RR is partially supported by a CNPq, Brazil, 
research grant 309158/2006-0. 

\end{document}